\documentstyle[12pt,aaspp4]{article}
\begin{document}
\parindent=1.0cm

\title{Stars, Star Clusters, and Dust in NGC 3077}

\author{T. J. Davidge \footnote[1]{Visiting Astronomer, Canada-France-Hawaii
Telescope, which is operated by the National Research Council of Canada,
the Centre National de la Recherche Scientifique, and the University of
Hawaii}\footnote[2]{This publication makes use of data products from the 
Two Micron All Sky Survey, which is a joint project of the University of 
Massachusetts and the Infrared Processing and Analysis Center/California 
Institute of Technology, funded by NASA and the NSF.}}

\affil{Canadian Gemini Office, Herzberg Institute of Astrophysics,
\\National Research Council of Canada, 5071 West Saanich Road,
\\Victoria, B.C. Canada V9E 2E7\\ {\it email: tim.davidge@nrc-cnrc.gc.ca}}

\begin{abstract}

	Images obtained with the CFHTIR camera on the Canada-France-Hawaii 
Telescope are used to investigate the near-infrared photometric properties of 
the star-forming M81 group galaxy NGC 3077. The spectral-energy distribution 
(SED) of the near-infrared light within 10 arcsec of the nucleus 
(1) is very different from that of `typical' dwarf ellipticals (dEs), blue 
compact dwarf galaxies, and HII/starburst galaxies, and (2) is consistent 
with the $2\mu$m light being dominated by hot young (log(t$_{yr}) < 6.8$) 
stars reddened by A$_V = 3 - 4$, with A$_V \geq 
8$ mag in some regions, including previously detected areas of CO 
emission. A population like that near the center of NGC 205 
likely contributes only a modest fraction of the light near $2\mu$m. 

	A number of candidate star clusters are detected in 
and around NGC 3077. These objects have near-infrared brightnesses and colors 
that are consistent with them being classical globular clusters and young star 
clusters. The specific frequency of globular clusters 
in NGC 3077 is estimated to be S$_N = 2.5$, 
which falls within the range of S$_N$ measured in nearby dEs. 
The candidate young clusters have photometric masses that are similar to 
those of compact young clusters in other active star-forming systems, and SEDs 
consistent with ages log(t$_{yr}) \leq 6.6$. Based on the masses and ages 
of the young clusters, it is estimated that the star formation rate in NGC 
3077 was at least $0.25 - 0.50$M$_{\odot}$ year$^{-1}$ during the past few 
million years. 

\end{abstract}


\section{INTRODUCTION}

	The origin of dwarf spheroidal (dSph) and dwarf elliptical (dE) 
galaxies, and their relation to dwarf irregular (dIrr) galaxies, has long been 
a matter of debate (e.g. Tajiri \& Kamaya 2002; Mayer et al. 2001; Silk, Wyse, 
\& Shields 1987; Thuan 1985; Lin \& Faber 1983). At issue is whether 
the morphological characteristics of these galaxies are primarily 
the result of (1) local conditions within the mini-halos from which 
they first collapsed, or (2) external factors, 
such as proximity to a much larger companion. 
Numerical simulations predict that the initial density 
and dark matter content are critical parameters for defining final morphology 
(e.g. Dekel \& Silk 1986; Ferrara \& Tolstoy 2000; Carraro et al. 2001), and 
the correlation between the chemical compositions, central surface brightness, 
and mass-to-light ratios of a wide range of dwarf galaxies is consistent with 
initial conditions playing a key role in dwarf galaxy evolution 
(Prada \& Burkert 2002). However, the occurence of 
morphological segregation in nearby groups (e.g. Karachentsev et al. 2002) and 
the relation between gas content and distance from nearby large 
companions in the Local Group (Blitz \& Robishaw 2000) indicate 
that environment also plays a key role in dwarf galaxy evolution. It is 
also clear that tidal interactions can profoundly affect the structural 
properties of dwarf galaxies in hierarchical systems, with the Sagitarrius 
dwarf being a prime example (Ibata, Gilmore, \& Irwin 1995).

	Studies of nearby galaxies and their companions will provide insights 
into dwarf galaxy evolution. The Milky-Way, M31, and M81 have similar masses 
(e.g. Kochanek 1996; Schroder et al. 2002; Perrett et al. 2002), morphologies, 
and environments, and yet have very different satellite systems. The 
companions of the Milky-Way include the dIrr Magellanic Clouds and a number of 
dSphs, while the brightest companions of M31 are the dE galaxies NGCs 
147, 185, and 205, and the compact elliptical galaxy M32; M31 also has a 
number of fainter dSph satellites. The companions of M81 show considerable 
diversity, consisting of dSphs, dEs, and dIrrs, and some of these 
show morphological peculiarities that are related to tidal interactions. 
Indeed, the M81 group appears to be undergoing significant evolution at the 
present day, making this the nearest laboratory for studying the 
effects of on-going tidal interactions on gas-rich dwarf galaxies.

	NGC 3077 is an actively star-forming member of the M81 group that is 
not easily placed within the Hubble sequence, although Price \& Gullixson 
(1989) conclude that there are similarities with the Local Group 
dE NGC 185 at red wavelengths. There is a  prominent central dust 
lane, and filamentary H$\alpha$ emission 
(Barbieri, Bertola, \& di Tullio 1974). The optical depth of the central 
dust lane is $\sim 0.5$, and this absorption obscures the isophotal center 
of the galaxy at visible wavelengths (Price \& Gullixson 1989). 
The central regions of NGC 3077 are dominated by hot stars (Benacchio \& 
Galletta 1981), and the star formation rate in the central 700 parsecs 
exceeds that in normal discs (e.g. Thronson, Wilton, \& Ksir 1991; Ott, 
Martin, \& Walter 2003). Martin (1998) identified a number of expanding gas 
shells in the ISM of NGC 3077. These structures have 
kinematic ages $\leq 10$ Myr, and spatial scales and expansion velocities 
that are indicative of energy input from a large number of SNe, thus confirming 
that there has been considerable recent star-forming activity. 

	NGC 3077 is not evolving in isolation. Ott et al. 
(2003) conclude that while the hot gas in NGC 3077 is confined 
at present, some of it may eventually escape into the M81 group intergalactic 
medium. An HI bridge connects M81 and NGC 3077 (van der Hulst 1979), while 
tidal spurs also link NGC 3077 to other members of the M81 group (e.g. Boyce 
et al. 2001). The interactions with M81 and its companions 
likely spurred the current star-forming episode in NGC 3077, and may also 
have triggered the formation of young compact star clusters in the M81 disk 
(Chandar et al. 2001).

	Studies of stars and star clusters in NGC 3077 will provide insight 
into the past history of this galaxy. The central regions of NGC 
3077 have yet to be resolved into stars. 
However, Sakai \& Madore (2001) and Karachentsev 
et al. (2002) resolved stars on the upper RGB in the outer regions of NGC 
3077, and the latter conclude that $\mu_0 = 27.91$, placing the system 
behind M81. The presence of RGB stars indicates that NGC 3077 
is not a recently formed tidal fragment, as has been
suggested for some other M81 companions (e.g. Karachentsev, Karachentseva, \& 
Boerngen 1985; Yun, Ho, \& Lo 1994; Boyce et al. 2001). 

	In the present study, deep $J, H,$ and $K'$ images 
obtained with the CFHTIR camera are used to probe the near-infrared 
spectral energy distribution (SED) and morphology of the central regions of 
NGC 3077, and search for star clusters. Observations of this galaxy 
in the infrared are of interest because the evolved cool stars that formed 
during intermediate and early epochs might be expected to dominate the light 
output at these wavelengths; knowledge of the spatial 
distribution of such stars may yield insight into the nature of NGC 
3077 prior to the most recent tidal interactions, and provide clues about its 
appearance after the current star-forming episode fades. Light at infrared 
wavelengths is also less affected by dust absorption than at visible 
wavelengths, making it easier to probe the heavily obscured central regions of 
NGC 3077, and detect bright star clusters.

	The paper is structured as follows. The observations and the data 
reduction procedures are described in \S 2. The infrared 
SED of integrated light in the central regions of NGC 3077 
and the isophotal properties of the galaxy are investigated in \S 3. 
Comparisons are also made with the Local Group dE galaxy 
NGC 205, which shows some similarities with NGC 3077, and may be in a more 
advanced evolutionary state (Davidge 1992). While the present data do not have 
the angular resolution needed to detect individual stars, a number of 
potential star clusters are identified, and the nature of these is investigated 
in \S 4. A summary and discussion of the results follows in \S 5.

\section{OBSERVATIONS AND REDUCTIONS}

	The data were recorded on June 4 2001 UT with the CFHTIR imager, which 
was mounted at the Cassegrain focus of the 3.6 metre Canada France Hawaii 
Telescope (CFHT). The detector in the CFHTIR is a $1024 \times 1024$ Hg:Cd:Te 
array. Each pixel subtends 0.2 arcsec on a side, so that a $3.4 \times 3.4$ 
arcmin field is imaged.

	Images of NGC 3077 were recorded through $J, H,$ and $K'$ filters 
with a total exposure time of 360 sec per filter. A four point square 
dither pattern was employed to assist with the identification 
of cosmic-rays and bad pixels, and to facilitate the 
construction of on-sky calibration frames. Faint standard 
stars from Hawarden et al. (2001) were also observed. The standard deviation 
of the $K$ zeropoints measured from individual stars is $\pm 0.04$ 
magnitudes. A comparison with sources in the 2MASS Point Source 
Catalogue confirms that the photometric calibration is reliable to 
within a few hundredths of a magnitude (\S 4.1).

	Each image was processed using the following sequence: (1) the 
subtraction of a master dark frame, which was constructed by combining dark 
frames taken with the same integration time as the science observation, 
(2) the division by a dome flat, which was constructed 
by taking the difference between images of a dome spot recorded with the dome 
lights on and off to remove thermal artifacts produced by warm objects along 
the line of sight, (3) the subtraction of the DC sky level, and 
(4) the subtraction of thermal artifacts and interference fringes from 
the on-sky images. The calibration images used in the last step were 
constructed from flat-fielded and sky-subtracted images of different fields, 
which were median-combined to reject stars and galaxies. The processed 
images were aligned to correct for the offsets in the dither pattern, and then 
median-combined to reject cosmic rays and bad pixels. The processed $K'$ image 
of NGC 3077, trimmed to the area of common integration time, and with the 
diffuse body of the galaxy, a template for which was obtained by median 
filtering the final processed image, subtracted out to allow sources near the 
galaxy center to be seen, is shown in Figure 1. Individual sources in the final 
combined images have a FWHM of 0.7 arcsec, and so the spatial resolution of 
these data, based on the Sakai \& Madore (1999) distance estimate, is roughly 
13 parsecs. 

\section{THE CENTRAL REGIONS OF NGC 3077}

\subsection{The Appearance and SED in the Near-Infrared}

	The central $12 \times 14$ arcsec of NGC 3077 in $K'$ is shown in the 
right hand panel of Figure 2, while a gray scale map of the $J-K$ color in 
this same region is shown in the left hand panel of this figure. 
The sky levels used to construct the color map were 
measured near the edges of the CFHTIR field in an effort to reduce 
contamination from the outer regions of the galaxy, and the maximum extent of 
the region shown in Figure 2 was selected to have a surface brightness 
that was high enough to be immune to the estimated uncertainties 
in the measured background level. Evidence that the photometric properties in 
this portion of NGC 3077 are not greatly affected by uncertainties in 
the sky level comes from the well-behaved nature 
of the $K$ surface brightness profile out to 20 arcsec radius, and the 
absence of systematic color gradients in this same region (\S 3.3).

	Images at visible wavelengths show dust lanes across the face of NGC 
3077 (Barbieri et al. 1974; Price \& Gullixson 1989), and so it is perhaps not 
surprising that there is structure in the color map, likely due to clumpiness 
in the dust distribution. There is good correspondence between the reddest 
areas in the color map and sources of CO emission: the central red 
feature in the color map corresponds to Walter et al. (2002) CO source R2, 
while the red area to the south is their CO source R1. The $J-K$ colors 
of R1 and R2 in 1 arcsec diameter apertures are 1.4 (R1) and 1.5 (R2), while 
$J - K \sim 1$ in the majority of pixels near the center of NGC 3077. 
If dust is the main cause of the color variations in Figure 2 then 
regions R1 and R2 are thus subjected to reddening in $J-K$ that is 
$0.4 - 0.5$ mag greater than the main body of the galaxy. 
In fact, the area near R1 and R2 is the site of 
relatively hard x-ray emission (Ott et al. 2003), which is likely a 
consequence of dust absorption. Finally, while R1 
appears as an elongated single source in the Walter et al. (2002) CO 
map, there are two distinct pockets evident in the $J-K$ map, suggesting 
that higher angular resolution CO observations will likely resolve R1 into 
(at least) two sources. 

	The $(J-H, H-K)$ two-color diagram (TCD) of individual pixels in the 
central $12 \times 14$ arcsec region of NGC 3077 is shown in Figure 3. The 
majority of pixels are concentrated near $J-H = 0.55$ 
and $H-K = 0.4$, and there is an extended plume 
of points that parallels the reddening vector, as expected if 
there are significant differences in the 
line of sight absorption near the center of NGC 3077. 
The overall length of the plume on the TCD parallel to the reddening 
vector suggests that differential reddening of size $\Delta$A$_V = 4$ 
magnitudes is present. Many of the pixels that are located 
above and to the right of the main body of points in the 
left hand panel of Figure 3 are in the vicinity of R1 and R2. Indeed, 
the colors of the CO source R1 are $J-H = 0.75$ and $H-K = 0.62$ 
within a 1 arcsec diameter aperture, while $J-H = 0.90$ 
and $H-K = 0.57$ for CO source R2. 

	To aid in the interpretation of the near-infrared SED of NGC 3077, 
stellar sequences from Bessell \& Brett (1988) are 
shown in Figure 3, along with the mean colors of blue compact dwarf galaxies 
(BCDGs), starburst/HII galaxies and Virgo cluster dEs based on data published 
by Thuan (1983), Hunt, Giovanardi, \& Helou (2002), and James (1994). 
It is evident that the near-infrared SED of NGC 3077 differs from that 
of stars and most other dwarf galaxies, including those that are currently 
forming stars. While the majority of pixels near the 
center of NGC 3077 have $J-H$ colors that do not differ from those of a 
typical Virgo cluster dE, these same pixels have $H-K$ 
colors that are $0.2 - 0.4$ mag redder than in a typical 
dE. The majority of pixels in the center of NGC 3077 also have colors that 
place them to the right of the BCDG and starburst/HII 
galaxy data on the TCD, although there is some overlap. Indeed, 
two of the BCDGs in the Thuan (1983) sample (Mrk 36 and Mrk 59) and two of 
the starburst/HII galaxies in the Hunt et al. (2002) sample (NGC 520 and NGC 
695) have integrated near-infrared colors that fall near the trend defined 
by the NGC 3077 data in Figure 3. Simple models of the near-infrared SED 
of NGC 3077 are explored in the next section.

\subsection{Modelling the Near-IR SED}

	As one of the nearest and best-studied dE galaxies, NGC 205 is a 
prime comparison object for NGC 3077, and the stellar contents 
of NGC 205 and NGC 3077 are similar in at least some respects. For example, 
there is a large population of cool AGB stars in NGC 205, indicating that there 
have been recent episodes of star formation (Richer, Crabtree, \& Pritchet 
1984; Davidge 1992, 2003; Lee 1996; Demers, Battinelli, \& Letarte 2003). 
In fact, the youngest stars in NGC 205 have ages log(t$_{yr}) \sim 8.0$ 
(e.g. Davidge 2003; Cappellari et al. 1999), which is comparable to the 
time since the last interaction between NGC 3077 and M81. 
Moreover, the young population in NGC 3077 is concentrated within the central 
700 parsecs (Thronson et al. 1991), while in NGC 205 the youngest 
stars are located within the central 600 parsecs (Davidge 2003).
Finally, the colors of bright AGB stars in NGC 205 suggest 
that the intermediate age population in this galaxy 
formed from gas with a near-solar metallicity (Davidge 1992). While the 
metallicity of NGC 3077 is uncertain, the emission spectrum suggests that gas 
in NGC 3077 may be as metal-rich as solar (Martin 1997). 

	$J, H,$ and $K'$ images of NGC 205 were obtained with the CFHTIR during 
the June 2001 observing run, and these were processed to simulate the 
appearance of this galaxy if viewed at the same distance and with the same 
angular resolution as NGC 3077. The NGC 205 images were reduced using the 
procedures described in \S 2, and the pixels in the results were block-averaged 
in $4 \times 4$ groups to simulate the loss of resolution due to the increase 
in distance. The block-averaged frames were then convolved 
with a $\sigma = 0.7$ arcsec Gaussian.

	The near-infrared TCD of pixels in the central $12 \times 14$ arcsec 
of the distance-processed NGC 205 image is shown in 
the right hand panel of Figure 3. The central regions of NGC 205 have an 
integrated SED that is similar to those of cool giants with modest amounts 
of reddening, and the Virgo cluster dEs studied by James (1994). For 
comparison, the near-infrared SED near the center of NGC 3077 is dominated by 
an extremely young population. To demonstrate this point, the locations of 
selected z=0.020, $\alpha = 2.35$, M$_{up} = 100$M$_{\odot}$ 
instaneous burst models from Leitherer et al. (1999) are shown in 
Figure 3. Solar metallicity models were selected for this comparison 
based on the strengths of emission lines in the spectrum of NGC 3077 
(Martin 1997). An extrapolation along the reddening vector from the main 
concentration of NGC 3077 data on the TCD passes close to the log(t$_{yr}$) = 
6.6 model; the log(t$_{yr}$) = 6.8 and 7.0 models have $H-K$ colors 
that are bluer than the SED of NGC 3077. The presence of a dominant 
population with an age log(t$_{yr}) < 7.0$ is consistent with (1) the ages 
of supershells in NGC 3077, which have log(t$_{yr}$) between 6.0 and 
7.0 (Ott et al. 2003), and (2) the large population of hot stars needed to 
produce the radio continuum flux measured by Meier, Turner, \& Beck (2001). 

	A system with a NGC 205-like SED can contribute only a modest 
fraction of the near-infrared light from the central regions of NGC 3077.
To demonstrate this point, the SEDs of the log(t) = 6.0 and 6.8 models 
were combined with the midpoint of the NGC 205 data distribution in the TCD in 
varying proportions and the results are shown in Figure 4. 
Models in which a log(t) = 6.0 population contributes at least $2 
\times$ the light from a population like that in NGC 205 in the $K-$band 
reproduce the unreddened near-infrared SED of NGC 3077 very well if the 
majority of pixels have A$_V$ between 3 and 4. The 
models involving the log(t) = 6.8 SED give a poorer match to the NGC 3077 
data for all possible contributions from a NGC 205-like SED. 
For example, a model made up exclusively of a population with log(t) = 6.8 
differs from the main concentration of NGC 3077 data on the TCD at roughly the 
$1-\sigma$ level, based on the estimated uncertainty in the photometric 
calibration, while a model in which a NGC 205-like system 
contributes 10\% of the near-infrared flux differs from 
the main concentration of NGC 3077 data at roughly the $2-\sigma$ level.

\subsection{The Near-Infrared Isophotal Properties of NGC 3077 and NGC 205}

	The $K'$ image in the right hand panel of Figure 2 shows little 
structure and is radially symmetric about the bright nucleus, which is 
roughly 1.8 arcsec north and east of the isophotal center of the galaxy, and 1 
arcsec to the north and east of the CO source R2. The general appearance of the 
$K'$ data suggests that non-uniformities in the dust 
absorption do not greatly affect the $K$ light profile of 
NGC 3077. This is not unexpected, given that A$_K = 0.11 \times$ A$_V$ 
(e.g. Rieke \& Lebofsky 1985), so that the $\Delta$A$_V = 4$ magnitude range in 
extinction, deduced from the distribution of points on the TCD in 
Figure 3, corresponds to only 0.4 mag in A$_K$.

	The structural properties of the central regions of NGC 3077 at 
near-infrared wavelengths have been investigated using the isophote-fitting 
routine Ellipse (Jedrzejewski 1987), and the radial behaviours of surface 
brightness, ellipticity, and position angle in the final $K'$ image 
are shown in Figure 5. Given that only the relatively high surface brightness 
central regions of the galaxy are considered it is perhaps not surprising 
that the estimated errors in the NGC 3077 surface 
brightness measurements due to uncertainties in the sky level are relatively 
modest, amounting to only 10\% at the largest radius considered. 

	The $K$ surface brightness profile of NGC 3077 can 
be matched with either an r$^{1/4}$ or exponential law outside of the seeing 
disk. The ellipticity increases, and the position angle drops, with increasing 
radius. Such radial changes in ellipticity and position angle are signatures 
of isophotal twisting, which simulations indicate could result from 
galaxy-galaxy interactions (e.g. Choi, Guhathakurta, \& Johnston 2002).

	Colors were also computed from the isophotal measurements. $J-H$ 
and $H-K$ do not change with radius out to 20 arcsec from the nucleus, with 
$\overline{J-H} = 0.55$ and $\overline{H-K} = 0.4$ in this region. 
Uncertainties in the sky background, coupled with the extended nature of 
NGC 3077, prevented the measurement of reliable colors at larger radii. 
However, NGC 3077 is in the 2MASS Extended Source Catalogue (Jarrett et al. 
2000), and the integrated colors measured from the 2MASS images are 
$J-H = 0.58$ and $H-K = 0.26$ within a radius of 109 arcsec, and $J-H 
= 0.57$ and $H-K = 0.25$ within a radius of 148 arcsec. These large radius 
2MASS measurements indicate that the $J-H$ and $H-K$ colors in the outer 
regions of NGC 3077 are in better agreement 
with those of BCDGs and Virgo cluster dEs than the central 
regions of the galaxy, suggesting that the young stellar content in NGC 
3077 is concentrated towards the center of the galaxy.

	The isophotal properties of NGC 205 were measured from the 
distance-shifted NGC 205 $K'$ images, and the results are shown in the right 
hand column of Figure 5. The $K'$ surface brightness profile of NGC 205 can be 
matched with either an r$^{1/4}$ or exponential law when $r > 1.5$ arcsec, in 
agreement with what was concluded by Kent (1987) from $r-$band data. The 
ellipticity and position angle of NGC 205 both change with radius in Figure 5, 
and similar trends are seen in Kent's (1987) $r-$band data.

	While the isophotal properties of NGC 205 and NGC 3077 
are not identical, there are similarities. In particular, although 
NGC 3077 has a higher mean surface brightness than NGC 205, the central 
$K$ surface brightnesses of both galaxies are roughly 1 mag arcsec$^{-2}$ 
brighter than would be inferred by extrapolating the surface brightness 
profile at larger radii to the galaxy center. Both galaxies also show radial 
changes in ellipticity and position angle. While NGC 205 shows a greater range 
in ellipticity than in NGC 3077, the amount of isophotal twisting, 
as measured from the change in position angle with radius outside of 
the seeing disk, is similar. Finally, after correcting for 
the presence of a young population, the $K-$band surface brightnesses of NGC 
3077 and NGC 205 are not greatly different. In \S 3.2 it was demonstrated that 
very young stars contribute at least twice as much as a NGC 205-like system to 
the integrated $K-$band light from NGC 3077. If the young stars are uniformly 
mixed with the NGC 205-like population near the center of NGC 3077, then the 
surface brightness of the underlying NGC 205-like component in NGC 3077 will be 
at least 1.2 magnitudes fainter in $K$ than in the composite system. Shifting 
the $K-$band surface brightness profile of NGC 3077 fainter by this amount, 
and brightening the results by 0.3 -- 0.4 magnitudes to correct for dust 
absorption in NGC 3077, yields a $K-$band surface brightness profile that is 
in much better agreement with that observed for NGC 205 than 
the original NGC 3077 profile. 

\section{STAR CLUSTERS IN NGC 3077}

	Studies of star clusters in NGC 3077 will 
provide insight into the evolution of the galaxy. 
The approximate faint limit of the current data is $K = 18$, and there are 
a number of sources brighter than this in the CFHTIR 
field that may be clusters belonging to NGC 3077. 
The locations, brightnesses, and colors of 
sources with $K \leq 18$ are summarized in Table 1. Crowding makes it 
difficult to resolve individual sources in the central 15 arcsec of NGC 3077 
and, while objects with $K \leq 18$ are present near the center of the galaxy, 
there is a chance that these may be blends of bright stars or clusters. 
Therefore, only objects with $K < 16$ within 15 arcsec of the nucleus 
are listed in Table 1.

	The $x,y$ co-ordinates in Table 1 are in pixel units, as 
measured by the DAOPHOT FIND routine. The origin is in the lower left hand 
corner of Figure 1, while the upper right hand corner is 
$(x,y) = (1000,1000)$. The columns labelled $\Delta\delta$ 
and $\Delta\alpha$ give offsets, in arcsec, from the isophotal 
center of NGC 3077 along the declination and right ascension axes, with 
positive offsets to the north and west.
The CFHTIR science field was not perfectly aligned 
along the cardinal axes, and the entries in the fourth and 
fifth columns of Table 1 have been adjusted for this. 
A scale of 0.211 arcsec pixel$^{-1}$ was adopted when computing these offsets.

	Some of the sources listed in Table 1 are in the 2MASS Point Source 
Catalogue (Cutri et al. 2003), and hence can be used to check the photometric 
calibration. There are 5 sources in common with the 2MASS survey that are 
brighter than $K = 16$ and are located more than 0.5 arcminutes from the center 
of NGC 3077. A comparison between the CFHTIR and 2MASS photometry gives 
mean differences, in the sense Davidge -- 2MASS, of $\Delta K = 
0.03 \pm 0.03$, $\Delta(J-H) = 0.05 \pm 0.07$, 
and $\Delta (H-K) = 0.01 \pm 0.04$, where the quoted
errors are the standard deviations about the mean 
difference. These mean differences are consistent with the uncertainties in 
the photometric calibration estimated from standard star measurements (\S 2).

	The objects in Table 1 have been classified as either 
foreground stars (FS), old globular clusters (OGC), or young clusters (YC) 
based on their near-infrared photometric properties. 
The near-infrared CMDs and TCD of these objects are shown in Figures 6 and 7; 
objects identified as foreground stars are not shown in Figure 7 to prevent 
cluttering this figure. The criteria used to make the identifications are 
discussed in the following sections. 

\subsection{Foreground Stars}

	Likely foreground stars can be identified using 
near-infrared colors and brightnesses. The majority of foreground 
stars should be on the lower main sequence and, 
unless they have very late spectral-types, 
will have $H-K < 0.3$, and $J-H < 0.7$ (e.g. Bessell \& Brett 
1988). Foreground stars will also have stellar SEDs and, 
since NGC 3077 is at a Galactic latitude $b \sim 40^o$, 
will likely have only modest amounts of reddening. 
Finally, some foreground stars may be too bright to be plausible members of 
NGC 3077. 

	There are eight sources with $K < 16.8$ that form a well-defined 
sequence in the CMDs in Figure 6, with $\overline{J-K} = 0.9$ and 
$\overline{H-K} = 0.2$. While not plotted in Figure 7, these objects also fall 
close to the stellar sequences on the TCD. Hence, these objects are identified 
in Table 1 as foreground stars.

\subsection{Globular Clusters and the Specific Cluster Frequency}

	Harris \& van den Bergh (1981) defined the specific frequency of 
globular clusters, $S_N$, as the number of clusters per 
unit galaxy luminosity, normalized to M$_B = -15$. Subsequent studies have 
demonstrated that there are systematic galaxy-to-galaxy differences 
in $S_N$, that are related, at least in part, to the structural properties 
of the host galaxy. In particular, S$_N$ in disk-dominated Local Group galaxies 
is an order of magnitude lower than in spheroidal 
systems (e.g. van den Bergh 1995). The Local Group dEs NGC 
147, 185, and 205 have $\overline{S_N} = 4.3 \pm 1.2$ (Harris 1991), which is 
similar to S$_N$ measured in Virgo and Fornax cluster dE's 
(Durrell et al. 1996; Miller et al. 1998), while Local Group dIrr 
galaxies have markedly lower S$_N$'s, with S$_N = 0.4 - 0.5$ in the SMC and 
LMC (Harris 1991). Differences of this nature may be due 
to the properties of the fragments from which the 
galaxies were assembled (Harris \& Harris 2002). 

	The globular cluster content of NGC 3077 may provide clues about the 
nature of the galaxy prior to the recent interactions. If NGC 3077 was 
originally a dE galaxy and has not lost a significant number of clusters due to 
tidal stripping, then it should have an entourage of some 30 globular clusters 
if M$_B = -17$ (Walter et al. 2002) and S$_N = 4$. On the other hand, if 
NGC 3077 were initially a dIrr with S$_N = 0.5$ then there should be 
at most 3 -- 4 clusters.

	The typical brightness of globular clusters in NGC 3077 can be 
estimated from the cluster content of other nearby galaxies. 
The peak luminosity of the globular cluster luminosity function 
(GCLF) remains constant to within a few tenths of a magnitude both in 
(e.g. Kavelaars \& Hanes 1997) and between (e.g. Kundu 
\& Whitmore 2001; Larsen et al. 2001; Harris 1991) galaxies. 
The M31 cluster system is the best studied extragalactic system, and 
the apparent magnitudes expected for globular clusters in NGC 3077 
can be estimated from the M31 cluster data. 
The brightest globular cluster in M31, 023--078, has $K =10.7$ 
(Barmby et al. 2000), which corresponds to M$_K = -13.8$, or $K = 14$ at the 
distance of NGC 3077. Hence, objects brighter than $K = 14$ near 
NGC 3077 are almost certainly {\it not} globular clusters. The brightest 
globular cluster in NGC 3077 will likely be much fainter than 023--078, 
as NGC 3077 is much less massive than M31.

	The peak of the M31 GCLF occurs near M$_K = -10$ (Barmby, Huchra, \& 
Brodie 2001), which corresponds roughly to $K = 18$ in NGC 3077. Consequently, 
globular clusters in NGC 3077 should occur in ever 
increasing numbers near the faint limit of the CFHTIR data. 
In fact, a population of objects with a broad range of colors is evident 
in the CMDs in Figure 6 when $K > 16.8$, and when placed on the TCD these 
objects have SEDs that are similar to those of classical globular 
clusters in M31, with only modest amounts of dust absorption. 
Twelve sources are identified as classical globular clusters in this way 
in Table 1. Some of these will almost certainly turn out to be 
foreground stars. Indeed, based on number counts between $K = 16$ and 
17, it can be anticipated that 3 -- 4 of the suspected classical globular 
clusters with $K$ between 17 and 18 are foreground stars.

	The candidate globular clusters in Table 1 sample only the 
upper half of the expected range of cluster brightnesses, and so a number 
of fainter globular cluster candidates wait to be discovered. If, as suggested 
by number counts of foreground stars with $K$ between 16 and 17, 8 
of the candidate clusters are confirmed as belonging to NGC 3077, and half 
of the cluster content has yet to be detected because it is below the detection 
limit of the CFHTIR data, which is near the approximate mid-point of 
the globular cluster luminosity distribution, then S$_N = 2.5$ in NGC 3077. 
For comparison, S$_N = 2.3 \pm 0.3$ in NGC 205 (Harris 1991). The specific 
frequency of globular clusters in NGC 3077 thus appears to overlap with that 
observed in dE galaxies.

\subsection{Young Clusters}

	The formation of compact star clusters is 
associated with large-scale star-forming episodes (e.g. Conti \& Vacca 1994; 
Whitmore et al. 1999), when massive molecular clouds are expected to be 
present (e.g. Harris \& Pudritz 1994). Given the level of 
star-forming activity, Chandar et al. (2001) suggest that massive 
young star clusters might be present in NGC 3077. If the 
clusters are sufficiently young, then they will have SEDs that differ markedly 
from those of old globular clusters, and so can be identified with the current 
data. In fact, the models of young populations from Leitherer et al. (1999), 
shown in Figure 3, predict that clusters with ages log(t$_{yr}) < 7$ 
have much redder $H-K$ colors than those of globular clusters, due to 
the contribution made by ionized gas to the SED of young systems. Fourteen 
objects with very red colors are seen in the CMDs in Figure 
6, and these are identified as compact young clusters in Table 1.

	When placed on the TCD in Figure 7, the candidate young clusters have 
SEDs that are similar to those of individual pixels near the center of NGC 
3077. In addition, the suspected young clusters form a sequence in the TCD that 
parallels the reddening vector, indicating that dust is likely 
the dominant source of dispersion in the near-infrared SEDs of these objects. 
Comparisons with the Leitherer et al. (1999) models in Figure 7 suggest 
that the candidate young clusters have ages log(t$_{yr}$) between 6.0 and 6.6, 
which is comparable to the ages of gas shells in NGC 3077 (Ott et al. 2003). 

	The line of sight extinction for each candidate young cluster was 
estimated by extrapolating along the reddening 
vector to a point midway between the log(t$_{yr}$) = 6.0 and 6.6 
models in Figure 7, and the results, along with the 
absolute brightness of each source, are summarized in Table 2. 
The brightest blue supergiants in galaxies have M$_K = -10$ 
(Rozanski \& Rowan-Robinson 1994), and the 
suspected young clusters identified here are brighter than this; hence, 
the sources listed in Table 2 are too bright to be individual massive stars. 

	The masses of the suspected young clusters were estimated using the 
M$_K$ values in Table 2 and the integrated $K-$band brightnesses predicted 
by the z=0.020 Leitherer et al. (1999) models with $\alpha = 
2.35$, M$_{low} = 1$M$_{\odot}$, and M$_{up} = 100$M$_{\odot}$. 
At a fixed mass, the integrated brightness of young systems is age-sensitive, 
and so masses were computed for assumed ages log(t$_{yr}$) = 
6.0 and 6.6. The results are shown in the last two 
columns of Table 2. The masses computed in this way are not 
sensitive to the adopted metallicity, but they do depend on the nature of 
the assumed initial mass function. For 
example, masses computed using the M$_{up} = 30$M$_{\odot}$ Leitherer et al. 
(1999) models would be 0.4 dex higher than those listed in Table 2. The mass 
estimates are also sensitive to the adopted low mass cut-off, and models 
that assume M$_{low} < 1$M$_{\odot}$ will give higher cluster masses 
than those listed in Table 2. Finally, some of the $K-$band light 
may also come from thermal emission from hot dust, which may occur in dense 
concentrations of hot stars (e.g. Hunt et al. 2002), and emission of this 
nature will cause the masses in Table 2 to be over-estimated. There is also 
a bias towards detecting clusters with thermal dust emission, as these objects 
will be brighter at 2$\mu$m than clusters with similar mass but lacking this 
light component.

	The majority of suspected young clusters have masses between 
log(M$_{\odot}$) = 4 and 5, with a mean log(M$_{\odot}$) = 4.5 if log(t$_{yr}$) 
= 6.0, and 4.8 if log(t$_{yr}$) = 6.6. The uncertainties discussed in the 
preceeding paragraph notwithstanding, the masses derived for the candidate 
young clusters in NGC 3077 are in good agreement with those seen in other 
star-forming regions, such as the circumnuclear environment of M83 (Harris et 
al. 2001). The most massive cluster is located close to the isophotal 
center of the galaxy. The nuclei of nucleated dE (dE,n) galaxies have globular 
cluster-like masses (e.g. Binggeli \& Cameron 1991), and the masses estimated 
for the nuclear cluster in NGC 3077 in Table 2
(logM = 5.8 and 6.1 for log(t$_{yr}$) = 6.0 and 6.6) fall within the mass 
range of Galactic globular clusters (Pryor \& Meylan 1993). 

\section{DISCUSSION AND SUMMARY}

\subsection{The Near-Infrared SED of NGC 3077}

	Sub-arcsec angular resolution $J, H,$ and $K'$ images obtained with 
the CFHTIR camera have been used to investigate the near-infrared photometric 
properties of the central regions of the M81 group galaxy NGC 3077. 
In \S 3 it was demonstrated that (1) the integrated near-infrared SED of 
the central regions of NGC 3077 differs from that of `typical' dE's, 
BCDG's and HII/starburst galaxies, and (2) 
the light near $2\mu$m is dominated by very young stars. 
The latter result is perhaps not surprising, as 
previous studies have found that the SFR in NGC 3077 is high, although there is 
significant scatter among the estimates. Thronson et al. (1991) estimate that 
the SFR is 0.06 M$_\odot$ year$^{-1}$ based on the integrated H$\alpha$ flux, 
and $\leq 0.25$ M$_\odot$ year$^{-1}$ from FIR emission. Walter et al. 
(2002) compute a SFR of 0.05 M$_\odot$ year$^{-1}$ based on
extinction-corrected H$\alpha$ measurements. Meier et al. 
(2001) estimate that the SFR is 0.4 M$_\odot$ year$^{-1}$ from the 2.6 mm 
continuum flux, while Ott et al. (2003) estimate that the SFR is 
0.6 M$_\odot$ year$^{-1}$ based on the energy needed to create super gas 
shells. As noted by Ott et al. (2003), the high SFR inferred in their study may 
suggest that the SFR has dropped recently. While the present-day SFR 
in NGC 3077 is markedly lower than in M82 (e.g. Ott et al. 2003), 
the efficiency with which stars form out of molecular gas in both galaxies 
is similar (Walter et al. 2002). 

	The detection of a population of 
suspected young star clusters, which have ages 
log(t$_{yr}$) between 6.0 and 6.8 and masses log(M$_{\odot}$) between 4.0 
and 5.0 (\S 4), is consistent with a very high recent SFR. The 
total integrated mass in the young clusters is $1 - 2 \times 10^6$ M$_{\odot}$. 
Given that the near-infrared SEDs of these clusters are suggestive of an age 
log(t$_{yr}) \leq 6.6$, then the SFR needed to produce these objects 
is $0.25 - 0.50$ M$_{\odot}$ year$^{-1}$, which falls within the range of 
estimates computed using other techniques. This is a lower limit to the 
total SFR, as it does not include stars that formed in clusters that have 
been disrupted, or clusters that are below the faint limit of these data.

	The near-infrared SED of NGC 3077 does not preclude a modest 
contribution from a system having an SED like that of 
NGC 205. The presence of such a population is consistent with the detection of 
RGB stars in NGC 3077 by Sakai \& Madore (2001) and Karachentsev 
et al. (2002), and the discovery of a healthy number of candidate old globular 
clusters (\S 4). Thus, NGC 3077 is not a recently formed system, as 
may be the case for some members of the M81 group 
(Yun et al. 1994, Boyce et al. 2001).

	The models used here to simulate the near-infrared SED of NGC 3077 
include contributions from stars and gas 
emission. However, Hunt et al. (2002) investigated the 
infrared photometric properties of star-forming galaxies, and concluded that 
thermal emission from hot dust may contribute significantly to the light from 
very active star-forming systems at wavelengths longward of $2\mu$m, and there 
are hints that emission from hot dust may contribute significantly to the 
infrared light from NGC 3077. One clue comes from the near-infrared SED. 
Thermal emission from dust with temperatures cooler than a few hundred K 
does not contribute significantly to light at wavelengths shortward of $2\mu$m, 
and so the $J-H$ color of a system with significant dust emission will be 
the same as from a galaxy lacking this emission, while the $H-K$ color may be 
much redder. It is evident from Figure 3 that the majority of pixels near 
the center of NGC 3077 have $J-H$ colors that are consistent with those of 
dEs and BCDGs, but have $H-K$ colors that are redder than in these systems, as 
expected if emission from hot dust is present. Emission from hot dust 
will likely be concentrated in the dense central star-forming 
regions of NGC 3077, where the radiation field is most intense. 
In \S 3.3 it was shown that the $J-H$ color measured near the center of NGC 
3077 is similar to that measured at larger radii from 2MASS survey data, while 
the central $H-K$ color is much redder than at large radii, and so the radial 
color profile is also consistent with emission from hot dust. Finally, when 
compared with other galaxies, NGC 3077 has a very low mass of cool gas (Stickel 
et al. 2000), as might be expected if a large fraction of the dust is heated by 
young stars. Observations of NGC 3077 in the $3 - 5\mu$m region will allow 
firmer constraints to be placed on any contribution made by 
thermal emission from hot dust. It is evident from Figure 4 that 
if significant emission from hot dust is present in NGC 3077 then the net 
result will be to allow for a larger contribution from a NGC 205-like SED. 

\subsection{The Evolution of NGC 3077}

	Tidal interactions are common events in the local 
Universe, as there are debris trails in the halos of the 
Milky-Way (e.g. Ibata et al. 1995), M31 (e.g. Ibata et 
al. 2001), and in nearby groups (e.g. Boyce et al. 2001).
Recognizing that tidal interactions can affect the structural 
characteristics of galaxies, and can also explain the morphology-density 
relation between dIrr and dSph systems, Mayer et al. (2001) 
suggested that the dSph and dE companions of the Milky-Way and M31 may have 
originally been gas-rich disky galaxies that were transformed by tidal 
interactions into spheroidal systems. Mayer et al. (2001) suggest 
that tides trigger bar instabilities that channel gas into the central regions 
of the progenitor, where star formation occurs. Feedback then heats 
the ISM, while the bar re-distributes angular momentum to the outer regions 
of the galaxy, which are subsequently stripped away. The bar eventually 
buckles, and the result is a system that has lost much of its angular momentum 
and has been transformed from a gas-rich disky to a gas-poor spheroidal 
morphology. Whether the final product is a dSph or a 
dE depends on the surface brightness of the progenitor, 
in the sense that dSphs evolve from low surface 
brightness dIrrs that experience multiple bursts of star formation, 
while dEs evolve from higher surface brightness dIrrs that experience 
one dominant, extended, star-forming episode that occurs over 
a $1 - 2$ Gyr period. Simulations indicate that the time scale 
for the transformation is a few Gyr, so there is a reasonable expectation 
of viewing this process at work in nearby galaxy groups. 

	While tidal interactions undoubtedly influence galaxy evolution, they 
do not provide a panacea for explaining dwarf galaxy morphology. Indeed, there 
are isolated dSph galaxies, such as Tucana, that likely have not been 
subjected to tidal interactions but still show remarkable similarities to dSphs 
in denser environments. The bar-driven transformation process described by 
Mayer et al. (2001) should have a major impact on radial population gradients, 
and yet the radial population behaviour of Tucana is similar to 
dSphs in denser environments (Harbeck et al. 2001). There are 
also galaxies in hierarchical systems that appear not to have been 
altered by tidal forces. In particular, the surface brightness profile and 
central black hole mass of M32, which is a galaxy that many have argued 
may be an extreme endpoint of tidal pruning (e.g. Faber 1973; Nieto 
1990), suggest that the structural characteristics of this galaxy were 
imprinted early on and have not since been greatly altered (Graham 2002).

	Tidal interactions are clearly affecting the properties of NGC 3077. 
Much of the atomic gas associated with NGC 3077 is in a tidal arm that is well 
offset from the main body of the galaxy (Walter et al. 2002; Yun et al. 1994), 
while there is a string of molecular complexes extending to the north and west 
of the center of NGC 3077 (Meier et al. 2001), the positioning of which 
may also be the result of tidal effects. 
While the interstellar medium of NGC 3077 is clearly being 
disrupted, when integrated over a large area, the HI mass to light ratio of 
NGC 3077 is more appropriate for a dIrr, rather than a dE (Walter et al. 
2002), as might be expected if NGC 3077 is undergoing a morphological 
transformation. Of course, it is possible that some of the gas 
currently associated with NGC 3077 may have been stripped from another galaxy. 

	Simulations discussed by Mayer et al. (2001) suggest that the time 
scale for bar evolution in tidally influenced gas-rich dwarf 
galaxies is on the order of a few Gyr, and so there is a reasonable 
expectation of observing barred tidally interacting dwarf galaxies. In fact, 
stars in the bar of the LMC have an age of $\sim 5$ Gyr (Smecker-Hane et al. 
2002), supporting the notion that the bars in dIrr galaxies in hierarchical 
systems can be stable against buckling for long periods of time. The 
timescale for bar disruption is much longer than the time 
since the last major encounter between M81, M82, and NGC 3077, so if a bar 
formed in NGC 3077 after the last encounter then it should still be present. 
This being said, the near-infrared images of NGC 3077 do not show 
evidence of a bar, and the peaky light profile of NGC 3077 (\S 3.3) is not 
consistent with a bar-dominated light distribution. 

	dIrr and dE galaxies have very different 
S$_N$'s (e.g. Harris 1991), and so the old globular cluster content of 
NGC 3077 may provide clues about the nature of the galaxy prior to its most 
recent encounter with M81. In particular, if NGC 3077 were a `typical' dE 
before encountering M81 then it should contain a rich population of classical 
globular clusters, while if it were recently a dIrr galaxy then it 
should contain only a modest population of such objects.
A caveat is that the young clusters that form during vigorous star-forming 
episodes can have globular cluster-like masses, as appears to be the case 
in NGC 3077 (\S 4.3). If these clusters are not disrupted then, when viewed in 
a few Gyr, the galaxy will contain an even larger globular cluster 
population, albeit spanning a range of ages. The S$_N$ may thus change with 
time over the course of the transformation process.

	Spectroscopic observations, which will yield line strengths and 
radial velocities, will be essential to distinguish between 
actual clusters and faint field stars in NGC 3077. The 
globular clusters in NGC 3077 will likely be more metal-poor than the 
surrounding field population (e.g. da Costa \& Mould 1988), and so 
will likely have weak absorption lines. 
Higher angular resolution images will also be useful for 
identifying clusters, as both globular clusters (Kundu \& Whitmore 
2001) and compact young clusters (Chandar et al. 2001; Harris et al. 2001) have 
characteristic sizes of a few parsecs, and so will have a non-stellar 
appearance when viewed with image qualities approaching 0.1 arcsec FWHM. 

	Although lacking spectra and high-resolution images, the present 
data still provide tantalizing hints into the nature of NGC 3077. In 
\S 4.2 it was estimated that S$_N = 2.5$, suggesting that the specific 
frequency of globular clusters in NGC 3077 is similar to that of NGC 205.
The specific globular cluster frequency is consistent with the 
structural characteristics of NGC 3077, which suggest that if star 
formation was terminated immediately then the galaxy would fade to become a 
dE,n. While the young central cluster in NGC 3077 is offset slightly from 
the isophotal center of the galaxy, this occurs in roughly 20\% of 
dE,n (e.g. Binggeli, Barazza, \& Jerjen 2000). 

\acknowledgements{Thanks are extended to the referee, Leslie Hunt, for 
providing comments that greatly improved the manuscript.}

\clearpage
\parindent=0.0cm

\begin{table*}
\begin{center}
\begin{tabular}{rrrccccrl}
\tableline\tableline
ID \# & $x$ & $y$ & $\Delta\alpha$ & $\Delta\delta$ & $K$ & $J-H$ & $H-K$ & Object Type\tablenotemark{a} \\
 & (pixels) & (pixels) & (arcsec) & (arcsec) & & & & \\
\tableline
1 & 31.5 & 461.8 & --100.1 & 0.0 & 17.987 & 0.689 & --0.018 & OGC \\
2 & 58.2 & 751.5 & --99.0 & $+61.0$ & 15.666 & 0.513 & 0.407 & FS \\
3 & 81.3 & 490.8 & --90.1 & $+6.3$ & 17.453 & 0.753 & 0.704 & YC \\
4 & 89.0 & 294.7 & --85.5 & --34.6 & 17.045 & 0.524 & 0.221 & OGC \\
5 & 122.7 & 674.8 & --84.2 & $+45.8$ & 15.027 & 0.612 & 0.272 & FS \\
6 & 154.3 & 239.2 & --70.9 & --45.4 & 16.709 & 0.992 & 0.794 & YC \\
7 & 157.4 & 140.4 & --68.7 & --66.1 & 16.901 & 0.756 & 0.662 & YC \\
8 & 180.8 & 786.2 & --73.7 & $+70.2$ & 16.676 & 0.647 & 0.206 & FS \\
9 & 229.8 & 534.8 & --59.5 & $+18.0$ & 16.993 & 0.867 & 0.643 & YC \\
10 & 281.0 & 212.0 & --43.8 & --49.1 & 16.608 & 0.693 & 0.629 & YC \\
11 & 282.1 & 555.8 & --48.8 & $+23.3$ & 15.636 & 0.562 & 0.231 & FS \\
12 & 290.4 & 564.2 & --47.2 & $+25.2$ & 16.670 & 0.679 & 0.646 & YC \\
13 & 310.9 & 312.7 & --39.0 & --27.5 & 17.755 & 0.933 & 0.411 & OGC \\
14 & 315.9 & 46.0 & --33.8 & --83.5 & 17.852 & 0.496 & 0.583 & YC \\
15 & 371.0 & 693.4 & --32.3 & $+53.6$ & 15.981 & 0.601 & 0.188 & FS \\
16 & 400.3 & 426.1 & --22.0 & --2.2 & 17.435 & 0.750 & 0.380 & OGC \\
17 & 411.7 & 443.2 & --19.8 & $+1.6$ & 16.871 & 0.776 & 0.308 & OGC \\
18 & 414.7 & 551.2 & --20.9 & $+24.3$ & 14.530 & 0.563 & 0.202 & FS \\
19 & 438.3 & 349.3 & --12.8 & --17.8 & 17.197 & 0.938 & 0.793 & YC \\
20 & 497.2 & 429.9 & --1.6 & 0.0 & 15.840 & 1.214 & 0.936 & YC \\
21 & 503.9 & 429.0 & 0.0 & 0.0 & 13.213 & 0.913 & 0.613 & YC \\
22 & 509.5 & 416.4 & $+1.2$ & --2.6 & 15.698 & 0.918 & 0.629 & YC \\
23 & 531.3 & 399.1 & $+6.0$ & --5.9 & 15.640 & 0.740 & 0.395 & YC \\
24 & 616.6 & 418.5 & $+23.7$ & --0.5 & 17.459 & 0.577 & 0.164 & OGC \\
25 & 669.1 & 701.1 & $+30.3$ & $+59.8$ & 12.575 & 0.644 & 0.237 & FS \\
26 & 761.8 & 940.1 & $+46.2$ & $+111.5$ & 16.476 & 0.605 & 0.222 & FS \\
27 & 763.1 & 585.6 & $+51.9$ & $+36.9$ & 16.986 & 0.582 & 0.135 & OGC \\
28 & 763.4 & 291.0 & $+56.5$ & --25.0 & 17.766 & 0.946 & 0.392 & OGC \\
29 & 771.8 & 347.2 & $+57.4$ & --13.1 & 17.098 & 0.575 & 0.173 & OGC \\
30 & 869.7 & 749.6 & $+71.8$ & $+73.1$ & 17.256 & 0.334 & 0.084 & OGC \\
31 & 908.2 & 779.4 & $+79.4$ & $+80.0$ & 17.488 & 0.685 & 0.198 & OGC \\
32 & 952.4 & 862.6 & $+87.5$ & $+98.2$ & 16.820 & 0.725 & 0.253 & OGC \\
33 & 964.2 & 157.7 & $+100.8$ & --50.0 & 17.042 & 0.864 & 0.636 & YC \\
\tableline
\end{tabular}
\end{center}
\caption{Objects With K $\leq 18$}
\tablenotetext{a}{FS = Foreground star; OGC = Old globular cluster; YC = Young cluster}
\end{table*}

\clearpage

\parindent=0.0cm

\begin{table*}
\begin{center}
\begin{tabular}{rcccc}
\tableline\tableline
ID \# & A$_V$ & M$_K^{Stellar}$ & log(Mass) & log(Mass) \\
 & & & (logt = 6.0) & (logt = 6.6) \\
\tableline
3 & 7.0 & -11.2 & 4.1 & 4.4 \\
6 & 9.1 & -12.2 & 4.5 & 4.8 \\
7 & 6.8 & -11.7 & 4.3 & 4.6 \\
9 & 7.4 & -11.7 & 4.3 & 4.6 \\
10 & 6.2 & -12.0 & 4.4 & 4.7 \\
12 & 6.2 & -11.9 & 4.3 & 4.6 \\
14 & 4.6 & -10.6 & 3.8 & 4.1 \\
19 & 8.7 & -11.7 & 4.3 & 4.6 \\
20 & 11.3 & -13.3 & 4.9 & 5.2 \\
21 & 7.7 & -15.5 & 5.8 & 6.1 \\
22 & 7.8 & -13.1 & 4.8 & 5.1 \\
23 & 5.5 & -12.9 & 4.7 & 5.0 \\
33 & 7.4 & -11.7 & 4.3 & 4.6 \\
\tableline
\end{tabular}
\end{center}
\caption{Intrinsic Parameters of the Candidate Young Clusters}
\end{table*}

\clearpage

\begin{center}
FIGURE CAPTIONS
\end{center}

\figcaption[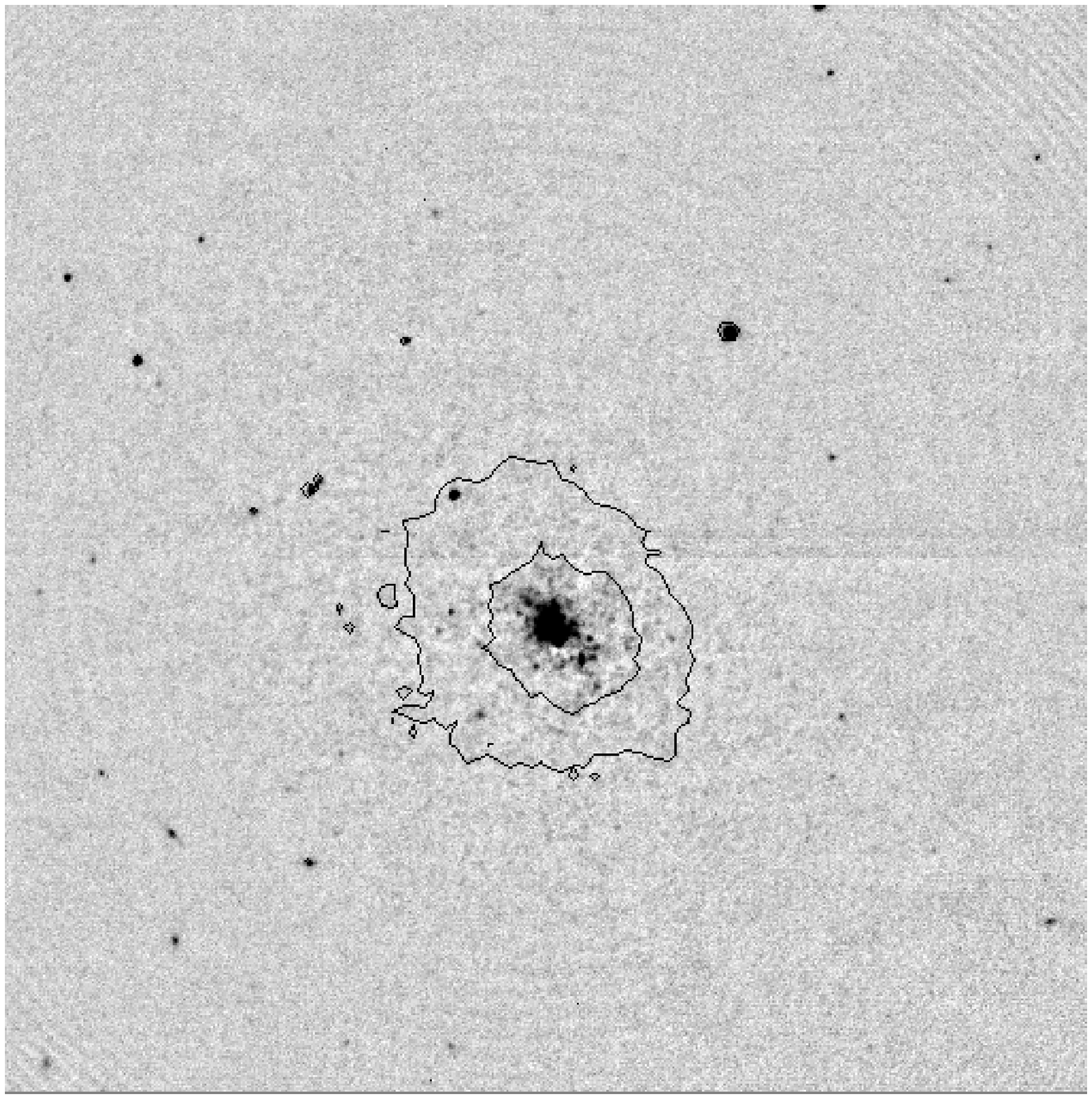]
{The $3.3 \times 3.3$ arcmin $K'$ image of NGC 3077. The diffuse light from the 
main body of the galaxy, a template for which was constructed by 
median-filtering the final processed $K'$ image, has been subtracted out to 
allow sources within a few tens of arcsec of the nucleus to be seen. North 
is at the top, and east is to the left. The image quality is FWHM = 0.7 arcsec. 
The contours show the $K = 20$ and 21 mag arcsec$^{-2}$ isophotes, 
measured from the final processed $K'$ image. The sources outside 
of the central regions of NGC 3077 are foreground stars and 
possible star clusters, and these are discussed in \S 4.} 

\figcaption[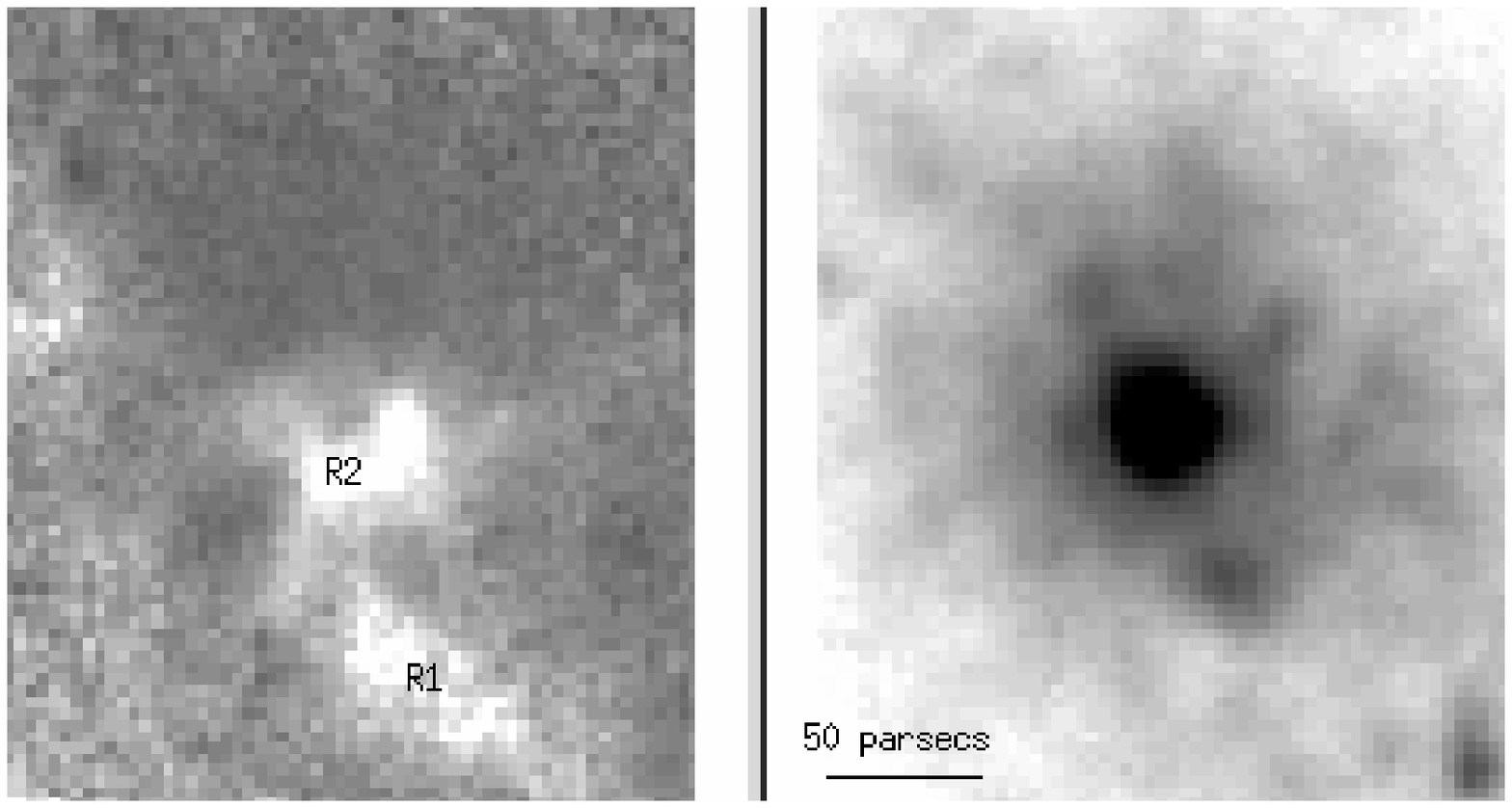]
{The central $12 \times 14$ arcsec of NGC 3077 as imaged in K' (right hand 
panel), and a gray scale map of the $J-K$ color in the same region (left hand 
panel). North is at the top, and east is to the left. The line in the lower 
left hand corner of the $K'$ image subtends a spatial distance of 50 parsecs 
based on the Sakai \& Madore (2001) distance modulus. Redder colors appear 
lighter in the $J-K$ image. The CO sources R1 and R2 (Walter et 
al. 2002) coincide with areas of relatively red color; 
CO source R3 is located just off of the left hand edge of the displayed field. 
The smooth appearance of the $K'$ image indicates that absorption from cool 
dust does not greatly affect the isophotal properties of NGC 3077 near $2\mu$m.}

\figcaption[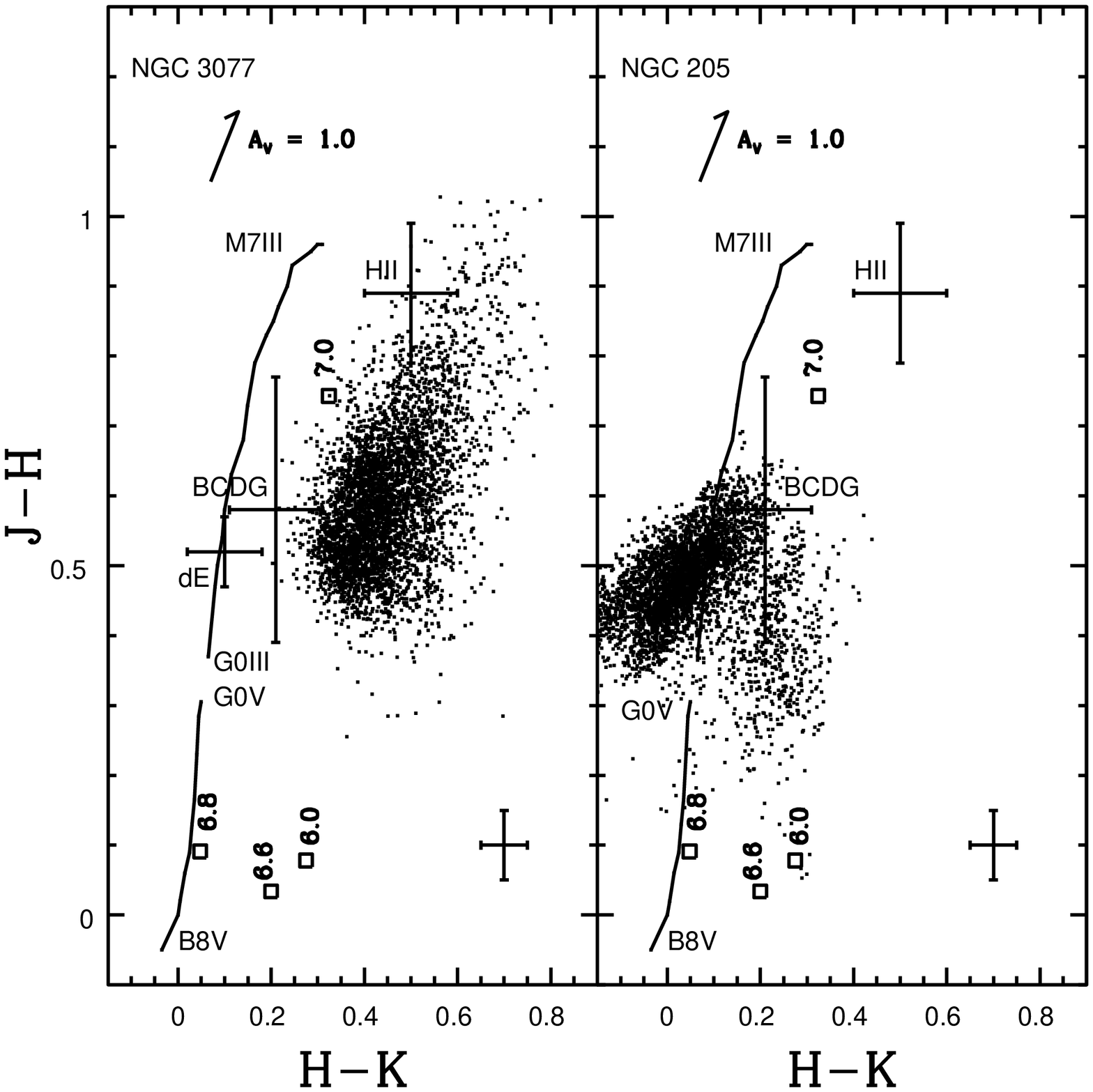]
{The $(J-H, H-K)$ TCD of pixels in the central $12 \times 14$ arcsec 
of NGC 3077 (left hand panel) and the corresponding 
region of NGC 205 (right hand panel). The error bars 
in the lower right hand corner of each panel show the uncertainties in the 
photometric calibration. The NGC 205 measurements were obtained from images 
that were processed to simulate the appearance of this galaxy if 
viewed at the same distance and with the same angular resolution as NGC 3077. 
The solar neightborhood giant and dwarf sequences from Tables II and III 
of Bessell \& Brett (1988) are shown, along with the mean locations 
of dEs, BCDGs, and starburst/HII galaxies in the samples observed by James 
(1994), Thuan (1983), and Hunt et al. (2002); the error bars show the standard 
deviation of the galaxy colors in each sample. The open squares show 
z=0.020 $\alpha=2.35$, M$_{up} = 100$ M$_{\odot}$ instantaneous burst models 
from Leitherer et al. (1999), labelled according to log(t$_{yr}$). The 
locus of the NGC 3077 data parallels the reddening vector, and the dispersion 
in the NGC 3077 data is suggestive of differential absorption of amplitude 
$\Delta$A$_V = 4$ magnitudes. An extrapolation along the reddening vector 
indicates that the light from the central regions of 
NGC 3077 has an unreddened SED that is similar to that 
of a simple stellar system with an age log(t$_{yr}) < 6.8$. For comparison, the 
majority of pixels near the center of NGC 205 have an SED similar to that 
of a `typical' Virgo cluster dE.} 

\figcaption[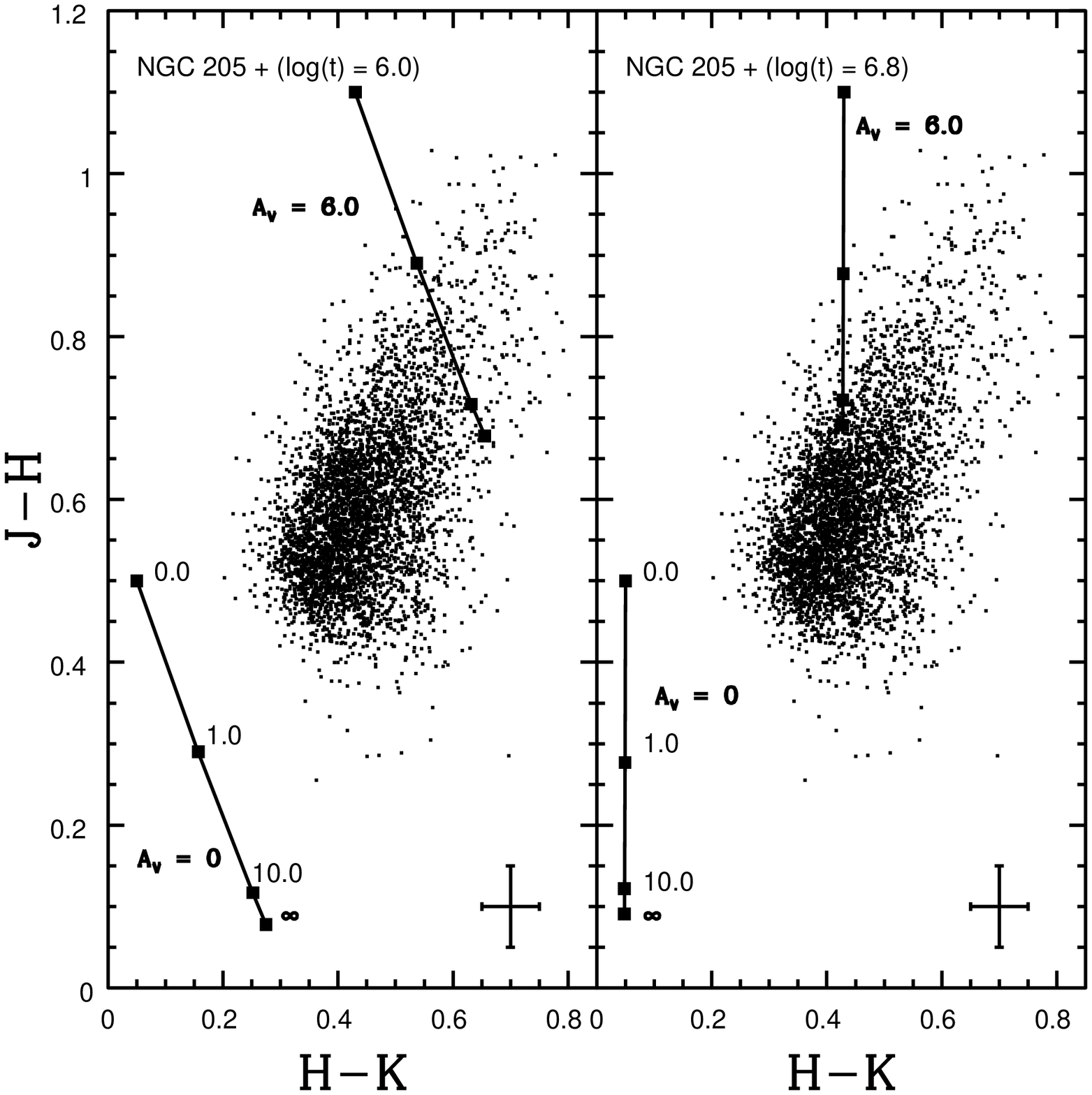]
{The $(J-H, H-K)$ TCD of pixels in the central $12 \times 14$ arcsec 
of NGC 3077. The error bars in the lower right hand corner of each panel show 
the uncertainties in the photometric calibration. The solid lines 
show the result of combining light from NGC 205 with the 
instaneous burst z = 0.020, $\alpha = 2.35$, and M$_{up} = 100$M$_{\odot}$ 
Leitherer et al. (1999) models with log(t$_{yr}$) = 6.0 (left hand panel) 
and 6.8 (right hand panel). Models with A$_V = 0$ magnitudes and A$_V = 
6.0$ magnitudes from a uniform foreground absorbing sheet 
are shown to demonstrate the effects of reddening. 
The ratio of light from the young population to an NGC 
205-like population in $K$ is indicated at various points along the 
model sequences. These models indicate that a significant fraction of 
the near-infrared light near the center of NGC 3077 must come from 
a relatively young population.}

\figcaption[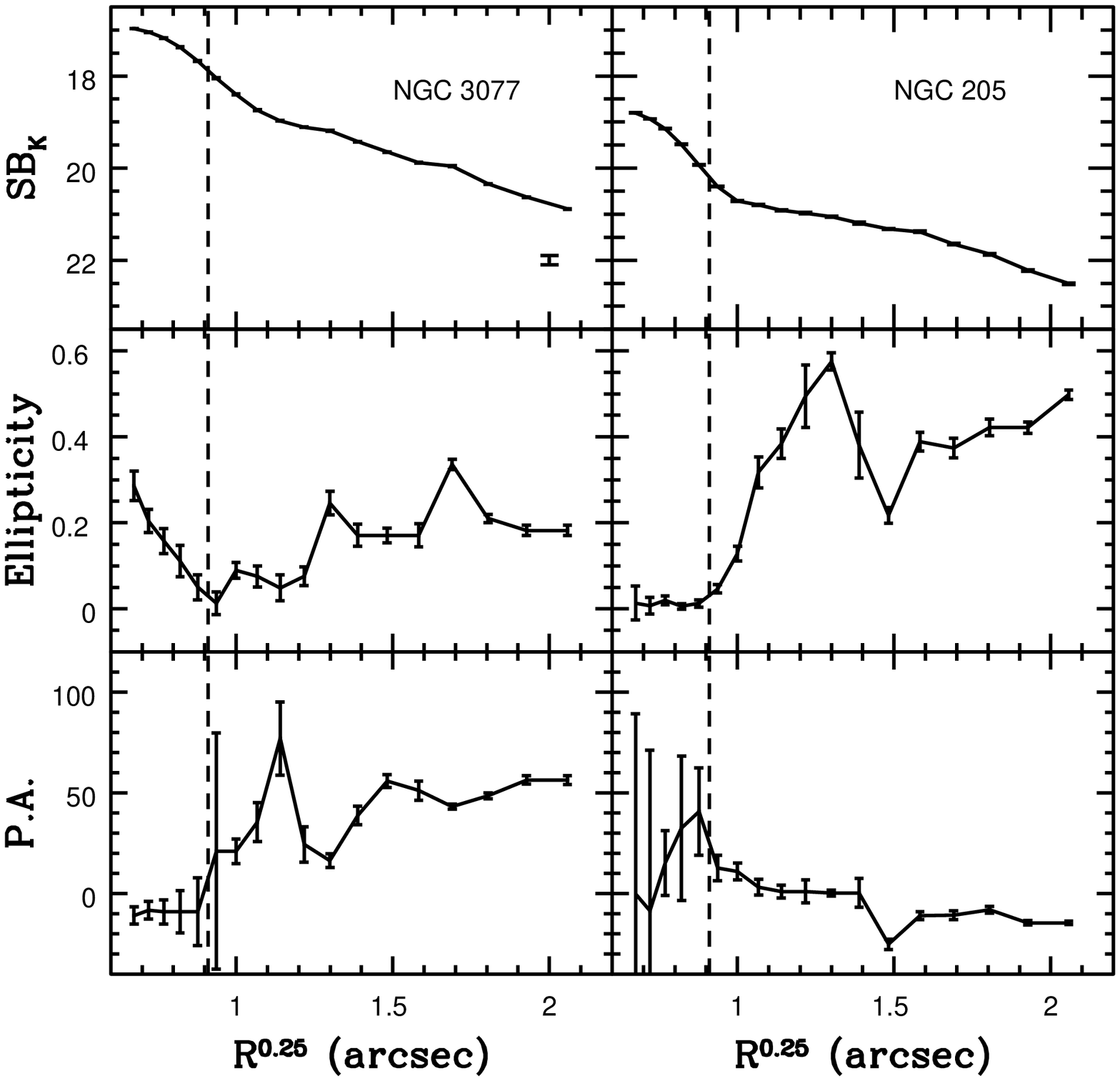]
{The surface brightness, ellipticity, and position angle profiles of 
NGC 3077 (left hand column) and NGC 205 (right hand column), as measured in 
$K'$. The NGC 205 measurements are from images of this galaxy that 
have been processed to simulate its appearance if viewed at the same distance 
and with the same angular resolution as NGC 3077. The 
dashed line marks the approximate extent of the seeing disk, and the 
measurements to the left of this line are blurred by seeing. 
In both galaxies the central $K$ surface brightness is roughly 
1 mag arcsec$^{-2}$ brighter than would be expected if the 
surface brightness profile at larger radii were extrapolated to smaller radii. 
The error bar in the lower right hand corner of the panel containing the 
NGC 3077 surface brightness measurements shows the estimated $\pm 0.1$ magnitude
($1-\sigma$) error in the last point due to uncertainties in the sky level.}

\figcaption[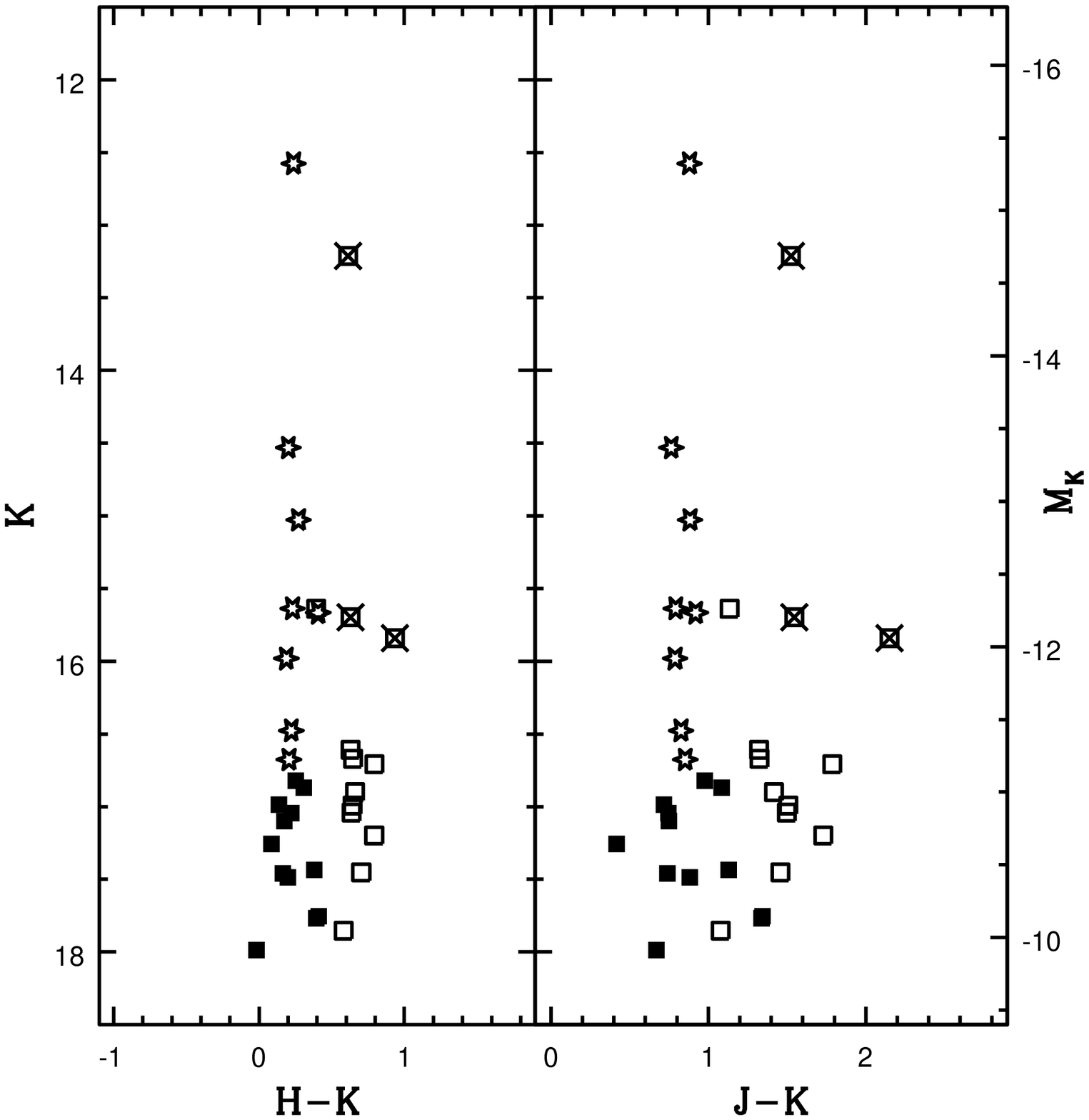]
{The $(K, H-K)$ and $(K, J-K)$ CMDs of foreground stars and possible star 
clusters in NGC 3077, identified using the photomeric criteria discussed in 
\S 4. Objects identified as classical globular clusters and young clusters are 
plotted as filled and open squares, respectively. Likely foreground stars 
are plotted as stars. Objects within 15 arcsec of the 
galaxy center, which are marked with a cross, have the reddest 
colors at a given $K$, as expected if the amount of dust absorption increases 
with decreasing distance from the center of NGC 3077. The absolute magnitude 
calibration on the right hand axis assumes a distance modulus $\mu_0 = 27.9$ 
(Sakai \& Madore 1999).}

\figcaption[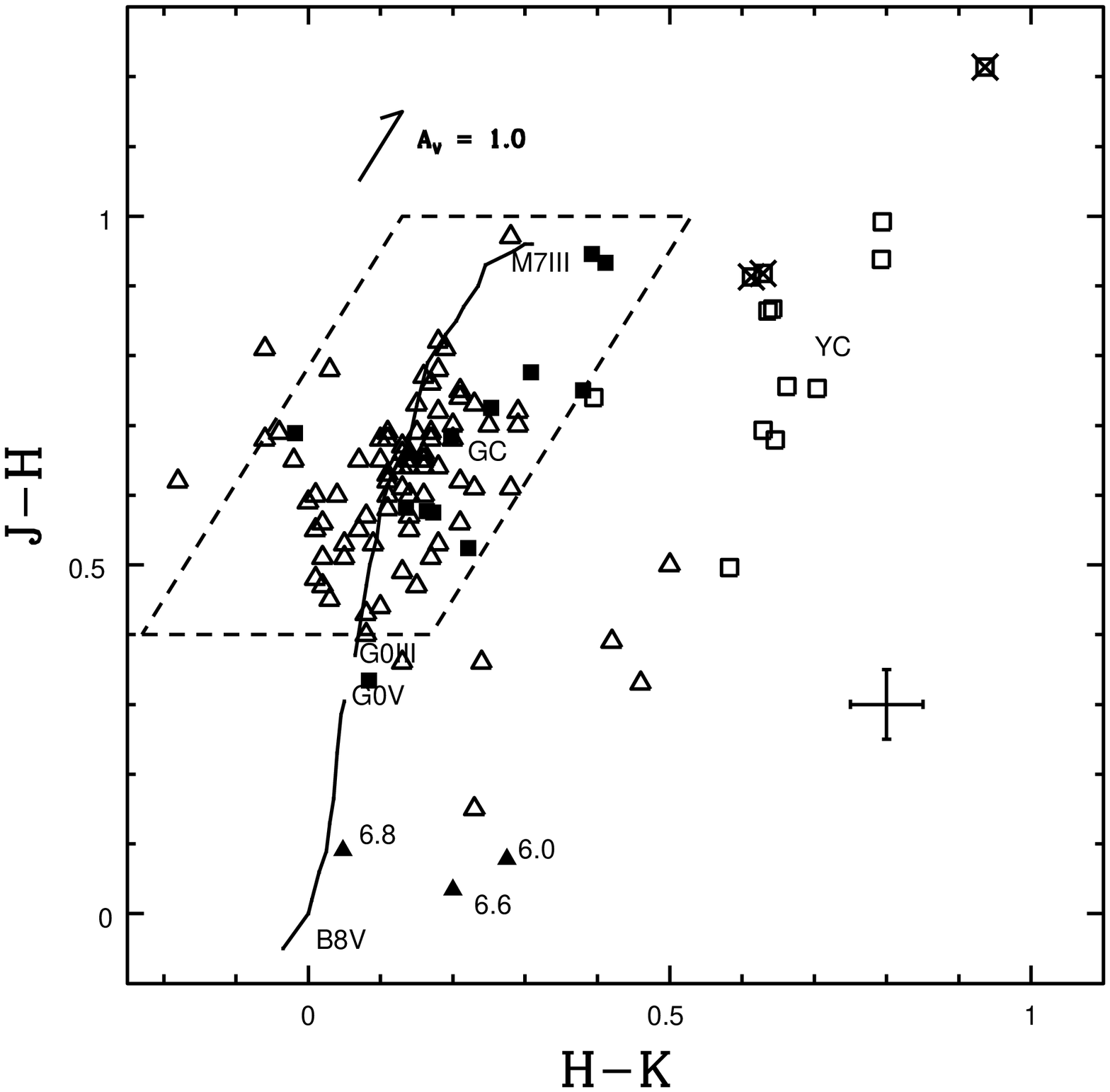]
{The $(H-K, J-H)$ TCD of possible star clusters in NGC 3077. The 
symbols are the same as in Figure 6. The error bar shows the estimated 
uncertainty in the photometric calibration. The solid lines show the sequences 
for solar neighborhood giants and dwarfs from Tables II and III of 
Bessell \& Brett (1988). The dashed lines mark the area 
containing the bulk of classical globular clusters in M31, 
which are plotted as open triangles using data 
from Tables 1 of Barmby et al. (2000, 2001), extended 
to account for possible line of sight extinction in NGC 
3077. The objects in the NGC 3077 data that fall within this area, and 
were not identified as foreground stars based on their colors and brightnesses 
(\S 4.1), are identified as globular clusters. The objects falling to the right 
of the globular cluster region are identified as possible young clusters. 
The filled triangles are z= 0.020 Leitherer et al. (1999) instanteous burst 
models with $\alpha = 2.35$ and M$_{up} = 100$M$_{\odot}$, which are labeled 
with log(t$_{yr}$). The majority of the candidate young clusters have SEDs 
consistent with reddened versions of the log(t$_{yr}) = 6.0$ and 6.6 models. 
The approximate midpoints of the regions containing the candidate globular 
clusters and young clusters are labelled with `GC' and `YC', respectively.} 
\end{document}